%% file: s2let_ridgelets_letter.tex
\documentclass[journal,a4paper]{IEEEtran}

\usepackage[]{graphicx}
\usepackage{url}
\usepackage{ifpdf}
\usepackage{hyperref}
\hypersetup{breaklinks,colorlinks,citecolor=blue}
\usepackage{cite}
\usepackage{amssymb}
\usepackage{amsmath}
\usepackage{mathtools}

\usepackage{subfigure}
\usepackage{color}

\include{macros}

\renewcommand{\eqn}[1]{Eq.~(#1)}

\renewcommand{\wav}{\ensuremath{\Psi}}

\renewcommand{\exp}[1]{\ensuremath{{\rm exp}{(#1)}}}

\renewcommand{\sshtcode}{{\sc ssht}}

\renewcommand{\stwoletcode}{{\sc s2let}}
\renewcommand{\healpix}{{\sc healpix}}
\newcommand{\fftwcode}{{\sc fftw}}

\begin{document}
%
\title{Scale-discretised ridgelet transform on the sphere}
%
%
%
\author{Jason~D.~McEwen and Matthew~A.~Price%
  \thanks{This work was supported by the Engineering and Physical
    Sciences Research Council (grant number
    EP/M011852/1)}
  \thanks{The authors are with the Mullard Space Science Laboratory
    (MSSL), University College London (UCL), Surrey RH5 6NT, UK.}%
  \thanks{E-mail: jason.mcewen@ucl.ac.uk }
}

\markboth{27th European Signal Processing Conference (EUSIPCO)}
{McEwen: Ridgelet transform on the sphere }
%



\maketitle

\begin{abstract}
  We revisit the spherical Radon transform, also called the
  Funk-Radon transform, viewing it as an axisymmetric convolution on
  the sphere.  Viewing the spherical Radon transform in this manner
  leads to a straightforward derivation of its spherical harmonic
  representation, from which we show the spherical Radon transform can
  be inverted exactly for signals exhibiting antipodal symmetry.  We
  then construct a spherical ridgelet transform by composing the
  spherical Radon and scale-discretised wavelet transforms on the
  sphere.  The resulting spherical ridgelet transform also admits
  exact inversion for antipodal signals.  The restriction to antipodal
  signals is expected since the spherical Radon and ridgelet
  transforms themselves result in signals that exhibit antipodal
  symmetry.  Our ridgelet transform is defined natively on the sphere,
  probes signal content globally along great circles, does not exhibit
  blocking artefacts, supports spin signals and exhibits an exact and explicit inverse transform.  No alternative ridgelet
  construction on the sphere satisfies all of these properties.  Our
  implementation of the spherical Radon and ridgelet transforms is
  made publicly available.  Finally, we illustrate the effectiveness
  of spherical ridgelets for diffusion magnetic resonance imaging of
  white matter fibers in the brain.
\end{abstract}

\begin{IEEEkeywords}
Harmonic analysis, spheres, spherical Radon transform, Funk Radon
transform, spherical wavelets, spherical ridgelets.
\end{IEEEkeywords}

%
\IEEEpeerreviewmaketitle

\section{Introduction}

\IEEEPARstart{W}{avelet} transforms on the sphere are becoming a
standard tool for the analysis of data acquired on a spherical
domain \cite{antoine:1999,
  antoine:1998, wiaux:2005, mcewen:2008:fsi, mcewen:szip,
  narcowich:2006, baldi:2009, marinucci:2008, wiaux:2007:sdw,
  leistedt:s2let_axisym, mcewen:2013:waveletsxv, sanz:2006,
  mcewen:2006:cswt2, starck:2006,
  simons:2011, simons:2011a, vielva:2004, mcewen:2005:ng,
  mcewen:2006:ng, mcewen:2008:ng, mcewen:2006:bianchi,
  mcewen:2006:isw, mcewen:2007:isw2, planck2013-p06, planck2013-p09, planck2013-p20}.

 Of particular note are discrete
wavelet frameworks on the sphere, which can support the exact
synthesis of signals from their wavelet coefficients in a stable
manner \cite{narcowich:2006, baldi:2009,
  marinucci:2008, wiaux:2007:sdw, leistedt:s2let_axisym, mcewen:2013:waveletsxv, starck:2006}.  Many of these frameworks have been extended to spin signal and signals on the three-ball
\cite{durastanti:2014, leistedt:flaglets, mcewen:flaglets_sampta, lanusse:2012}.

However, the effectiveness of wavelets on the sphere is limited for highly anisotropic signal content.
Directional scale-discretised wavelets on the sphere
\cite{wiaux:2007:sdw, leistedt:s2let_axisym, mcewen:2013:waveletsxv,
  mcewen:s2let_spin, mcewen:s2let_localisation}
go some way to addressing this shortcoming, however geometric
properties of structures are not exploited.
In Euclidean space, alternative transforms such as ridgelets and
curvelets have been devised for such a purpose
\cite{candes:1999:ridgelets, candes:1999:curvelets, candes:2005a,
  candes:2005b}, which in turn (may) rely on the Radon transform
\cite{radon:1917, radon:1986}.

The spherical Radon transform, also called the Funk-Radon transform,
is constructed from the integration of a signal along great circles
\cite{funk:1916}.
In this letter we present a novel take on the spherical Radon
transform, viewing it as a convolution with a kernel defined by a
Dirac delta function in colatitude, such that it is non-zero along
the equatorial great circle only.  Viewing the spherical Radon
transform in this manner helps to aid intuition, which leads to a
straightforward derivation of its harmonic action (presented
previously \cite{anderson:2005, descoteaux:2007, hess:2006}
in an alternative manner). In addition, we show that
inversion of the spherical Radon transform is well-posed for
signals that exhibit antipodal symmetry, \textit{e.g.} in MRI
analysis.  While techniques that attempt to invert the spherical
Radon transform are typically approximate
\cite{andersson:1988, klein:2003, louis:2011, regnier:2013},
our inversion is exact and explicit.

First-generation ridgelets and curvelets were constructed on the
sphere in \cite{starck:2006}. However these wavelets are constructed
by performing ridgelet and curvelet transforms of the twelve base-resolution
faces of the \healpix $\!$ pixelisation of the sphere \cite{gorski:2005} and so do
not live natively on the sphere, do not probe signal content along great circles,
and may result in blocking artefacts, as ackowledged in \cite{starck:2006}.
Second-generation curvelets have recently been developed \cite{chan:s2let_curvelets}
which live natively on the sphere, exhibit the parabolic scaling typical of curvelets,
and do not suffer from blocking artefacts.

An alternative ridgelet transform on the sphere has been constructed
in \cite{michailovich:2010a}.  This construction lives natively on the
sphere, probes signal content along great circles and does not exhibit
any blocking artefacts. The ridgelet transform is constructed from a
standard spherical Radon transform, followed by a wavelet transform
on the sphere.  Although this construction has already been
demonstrated to be of considerable practical use
\cite{michailovich:2010a, michailovich:2010b}, the forward
ridgelet transform is approximated in an iterative manner by an orthogonal
matching pursuit algorithm and an explicit inversion is not given \cite{michailovich:2010a}.

In this letter we develop a second-generation ridgelet transform on the sphere that
exhibits all of the desirable properties of the construction of
\cite{michailovich:2010a} and exhibits an explicit forward and inverse transform that can be computed efficiently and exactly for signals
exhibiting antipodal symmetry.  Moreover, our construction supports spin signals.

The letter is structured as follows. First, we present a novel
take on the spherical Radon transform in \sectn{\ref{sec:radon}}, viewing
it as a convolution on the sphere, which leads to a straightforward derivation
of its harmonic action.  The spherical ridgelet transform is presented
in \sectn{\ref{sec:ridgelet}}. The numerical
implementation of our ridgelet transform is presented and evaluated in
\sectn{\ref{sec:evaluation}} and an illustrative application to
diffusion MRI is presented in \sectn{\ref{sec:illustration}}.
Concluding remarks are made in \sectn{\ref{sec:conclusions}}.

\section{Spherical Radon Transform}
\label{sec:radon}

We present a novel take on the well-known spherical Radon transform,
viewing it as an axisymmetric convolution, which leads to a
straightforward derivation of its harmonic action.

\subsection{Axisymmetric convolution}

The  axisymmetric convolution $\convaxisym$ of a function
$\fs \in \ltwo(\sphere)$ with an axisymmetric kernel
${}_\spin h \in \ltwo(\sphere)$ is defined by
\begin{align}
  \label{eqn:axisymmetric_convolution_spatial}
  ( \fs \convaxisym & {}_\spin h) (\sas)
  \equiv \innerp{\fs}{\rotarg{(\sas)} {}_\spin h} \nonumber \\
  &= \int_\sphere \dmu{\saa\p,\sab\p} \:
  \f(\saa\p,\sab\p) \: \bigl(\rotarg{(\sas)}{}_\spin h\bigr)^\cconj(\saa\p,\sab\p)
  \spcend ,
\end{align}
where we adopt the shorthand notation for the axisymmetric spherical
rotation operator $\rotarg{(\eulb,\eula)} \equiv \rotarg{(\eula,\eulb,0)} \in \sothree$ parameterised by the
Euler angles $(\euls)$. Axisymmetric convolution may be expressed
by its harmonic expansion:
\begin{equation}
  \label{eqn:axisymmetric_convolution_harmonic}
( \fs \convaxisym {}_\spin h) (\sas) = \sumlm
  \sqrt{\frac{4\pi}{2\el+1}} \:
  \sshc{\f}{\el}{\m}{\spin} \:
  \sshcc{h}{\el}{0}{\spin} \:
  \shfarg{\el}{\m}{\sas}
  \spcend ,
\end{equation}
for spin harmonic coefficients $\fslm = \innerp{\fs}{\sshf{\el}{\m}{\spin}}$ and $\sshc{h}{\el}{0}{\spin} \kron{\m}{0} = \innerp{{}_\spin
  h}{\sshf{\el}{\m}{\spin}}$. Notice that although two spin functions are convolved, the resultant
$( \fs \convaxisym {}_\spin h)$ is a scalar ($\spin=0$) function on
the sphere \cite{mcewen:s2let_spin, mcewen:s2let_localisation}.

\subsection{Forward transform}

The spherical Radon transform, also known as the Funk-Radon transform,
is given by \cite{funk:1916}
\begin{equation}
  (\sradon \fs) (\sas)
  \equiv \int_\sphere \dmu{\saa\p,\sab\p} \: \fs(\saa\p,\sab\p) \:
  \delta(  \vect{\hat{\sa}\p} \cdot \vect{\hat{\sa}})
  \spcend ,
\end{equation}
where $\vect{\hat{\sa}}$ and $\vect{\hat{\sa}\p}$ denote the Cartesian vector
corresponding to angular coordinates $\sa=(\sas)$ and
$\sa\p=(\saa\p,\sab\p)$, respectively.
In words, the spherical Radon transform is the collection of line
integrals of \fs\ along great circles with poles at $\sa=(\sas)$,
projected onto the point defined by the poles of the great circles.

By defining the Funk-Radon kernel
$\sradondelta(\sas) \equiv \delta(\saa-\pi/2)$, the spherical Radon
transform $(\sradon \fs) (\sas)$ may be expressed as an axisymmetric convolution by
\begin{align}
  ( \fs \convaxisym \sradondelta) (\sas) \spcend
  &= \int_\sphere \dmu{\saa\p,\sab\p} \: \fs(\saa\p,\sab\p) \:
    (\rotarg{(\sas)} \sradondelta)(\saa\p,\sab\p) \nonumber.
\end{align}
Consequently, by noting
\eqn{\ref{eqn:axisymmetric_convolution_harmonic}}, the spherical Radon
transform can be expressed in harmonic space by
\begin{equation}
  \shc{( \sradon \fs)}{\el}{\m}
  =
  \shc{( \fs \convaxisym \sradondelta)}{\el}{\m}
  =
  \sqrt{\frac{4\pi}{2\el+1}} \:
  \sshc{\f}{\el}{\m}{\spin} \:
  \sshcc{\sradondelta}{\el}{0}{\spin}
  \spcend ,
\end{equation}
where the harmonic coefficients of the Funk-Radon kernel read
\begin{equation}
  \sshc{\sradondelta}{\el}{\m}{\spin} =
  (-1)^\spin
  \sqrt{\pi(2\el+1)} \:
  \sqrt{\frac{(\el-\spin)!}{(\el+\spin)!}} \:
  \aleg{\el}{\spin}{0} \:
  \kron{\m}{0}
  \spcend \label{eqn:funk_radon_kernel_harmonic}.
\end{equation}

Viewing the Funk-Radon transform as an axisymmetric convolution
allows us to derive its harmonic representation in a straightforward manner as
\begin{equation}
  \label{eqn:funk_radon_harmonic}
  \shc{( \sradon \fs)}{\el}{\m}
  =
  2 \pi \:
  (-1)^\spin
  \sqrt{\frac{(\el-\spin)!}{(\el+\spin)!}} \:
  \aleg{\el}{\spin}{0} \:
  \sshc{\f}{\el}{\m}{\spin}
  \spcend .
\end{equation}

\subsection{Inverse transform}

An inverse function to \eqn{\ref{eqn:funk_radon_harmonic}} exists
if the associated Legendre functions are well-behaved at the origin.
It can be shown that
$\aleg{\el}{\spin}{0} = \order(\ell^{-1/2})$ as
$\ell \rightarrow \infty$, for $\spin \ll \ell$ (which is typically
the case in practice) and for $\el+\spin$ even,  while for $\el+\spin$ odd,
$\aleg{\el}{\spin}{0}=0$. Concequently, the spherical Radon
transform of signals with non-zero harmonic coefficients for $\el+\spin$
even only, can be inverted by
\begin{equation}
  \label{eqn:funk_radon_inverse_harmonic}
  \shc{\fs}{\el}{\m}
  =
  \shc{( \sradon^{-1} \sradon \fs)}{\el}{\m}
  \equiv
  \frac{  \shc{( \sradon \fs)}{\el}{\m} }
  {2 \pi \:
  (-1)^\spin
  \sqrt{\frac{(\el-\spin)!}{(\el+\spin)!}} \:
  \aleg{\el}{\spin}{0} }
  \spcend .
\end{equation}

For scalar signals, the restriction to signals with harmonic
coefficients non-zero for even $\el$ only corresponds to signals with
antipodal symmetry -- unsurpisingly, as the forward spherical Radon
transform necessarily produces antipodal signals.
In practice, inversion can be performed accurately up to very high $\el$.

\subsection{Properties}

We conclude our discussion of the spherical Radon transform by noting
two important properties.

\subsubsection{Shift invariance}

The spherical Radon transform is shift invariant, such that
\begin{equation}
  \bigl(\sradon \: \rotarg{(\euls)} \: \fs \bigr) (\sas)
  =
  \bigl(\rotarg{(\euls)} \: \sradon \: \fs \bigr) (\sas)
  \spcend .
\end{equation}

\subsubsection{Eigenfunctions and eigenvalues}

By considering the spherical Radon transform of the spin spherical
harmonics $\sshf{\el}{\m}{\spin}$, we see from \eqn{\ref{eqn:funk_radon_harmonic}} that
\begin{equation}
  (\sradon \sshf{\el}{\m}{\spin})(\sas)
  =
  {}_\spin \lambda_\el
  \:
  \sshf{\el}{\m}{\spin}(\sas)
  \spcend ,
\end{equation}
The spin spherical harmonics are therefore the eigenfunctions of the
spherical Radon transform, with corresponding eigenvalues
${}_\spin \lambda_\el =
2 \pi \:
(-1)^\spin
\sqrt{\frac{(\el-\spin)!}{(\el+\spin)!}} \:
\aleg{\el}{\spin}{0}$.

\section{Spherical Ridgelet Transform}
\label{sec:ridgelet}

We present a novel spherical ridgelet transform on the sphere by
composing the spherical Radon transform and the scale-discretised
wavelet transform.  Our construction permits an explicit inverse
transform to synthesise antipodal signals from their ridgelet
coefficients exactly and satisfies a number of additional desirable
properties. For a complete
review of scale-discretized wavelets see \cite{wiaux:2007:sdw,
leistedt:s2let_axisym, mcewen:2013:waveletsxv, mcewen:s2let_spin,
mcewen:s2let_localisation}.

\subsection{Ridgelet analysis and synthesis}

We define the ridgelet transform on the sphere by the axisymmetric
convolution with the ridgelet $\srwav^{(\wscale)} \in \ltwo(\sphere)$:
\begin{align}
  \label{eqn:ridgelet_analysis}
  \rwcoeff & {}^{\srwav^{(\wscale)}} (\sas)
  \equiv
  (\rwavop^{\srwav^{(\wscale)}} \fs)(\sas)
  \equiv ( \fs \convaxisym \srwav^{(\wscale)}) (\sas) \nonumber \\
  &= \int_\sphere \dmu{\saa\p,\sab\p} \: \fs(\saa\p,\sab\p) \:
    (\rotarg{(\sas)} \srwav^{(\wscale)})(\saa\p,\sab\p)
  \spcend ,
\end{align}
with ridgelet coefficients
$\rwcoeff^{\srwav^{(\wscale)}} \in \ltwo(\sphere)$ defined on the
sphere.

\begin{figure}
  \centering
  \subfigure[Colour plot for $\wscale=3$]{\includegraphics[trim=40mm 25mm 40mm 20mm,clip=true,width=.40\linewidth]{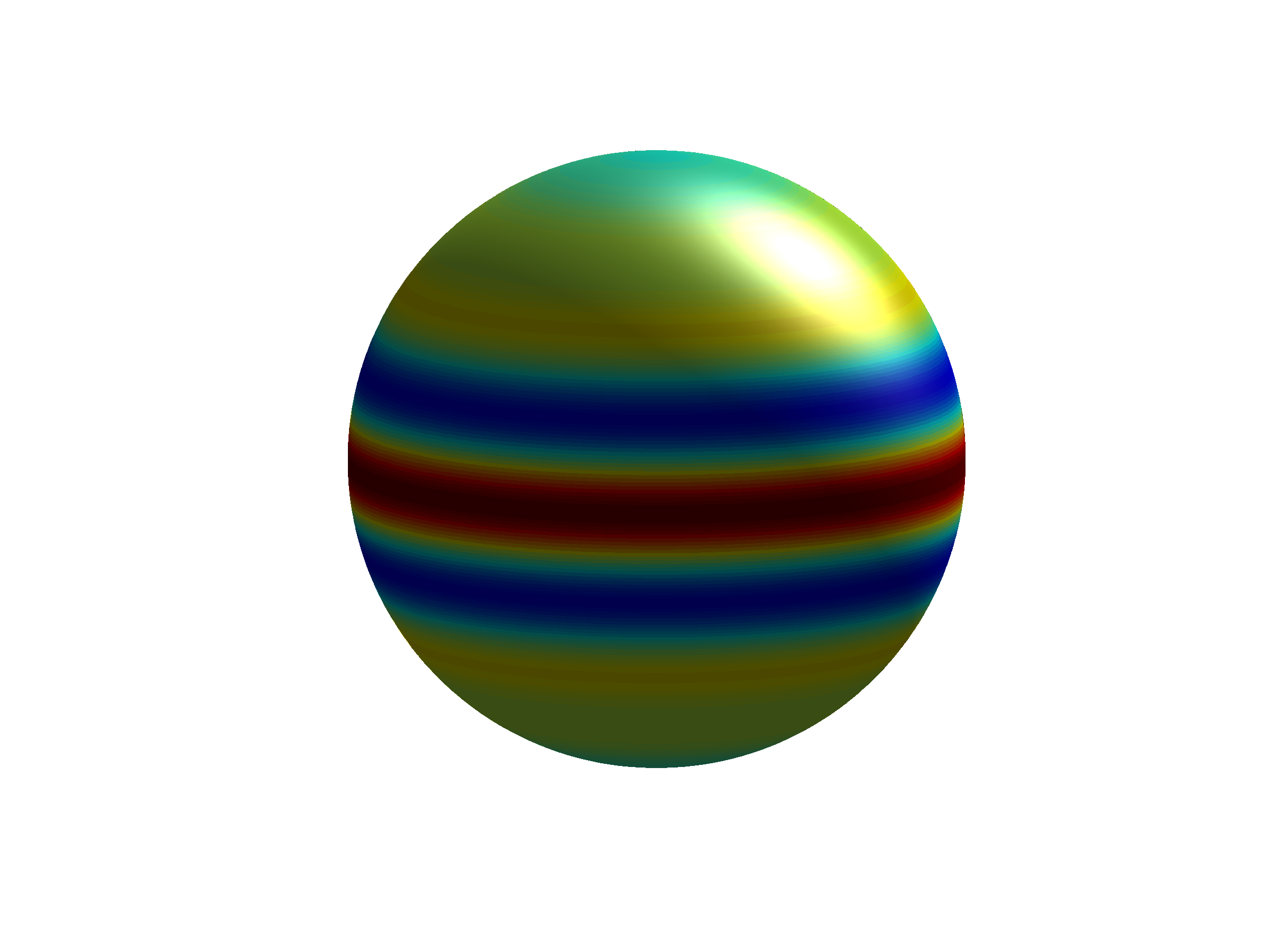}}\quad
  \subfigure[Colour plot for $\wscale=4$]{\includegraphics[trim=40mm 25mm 40mm 20mm,clip=true,width=.40\linewidth]{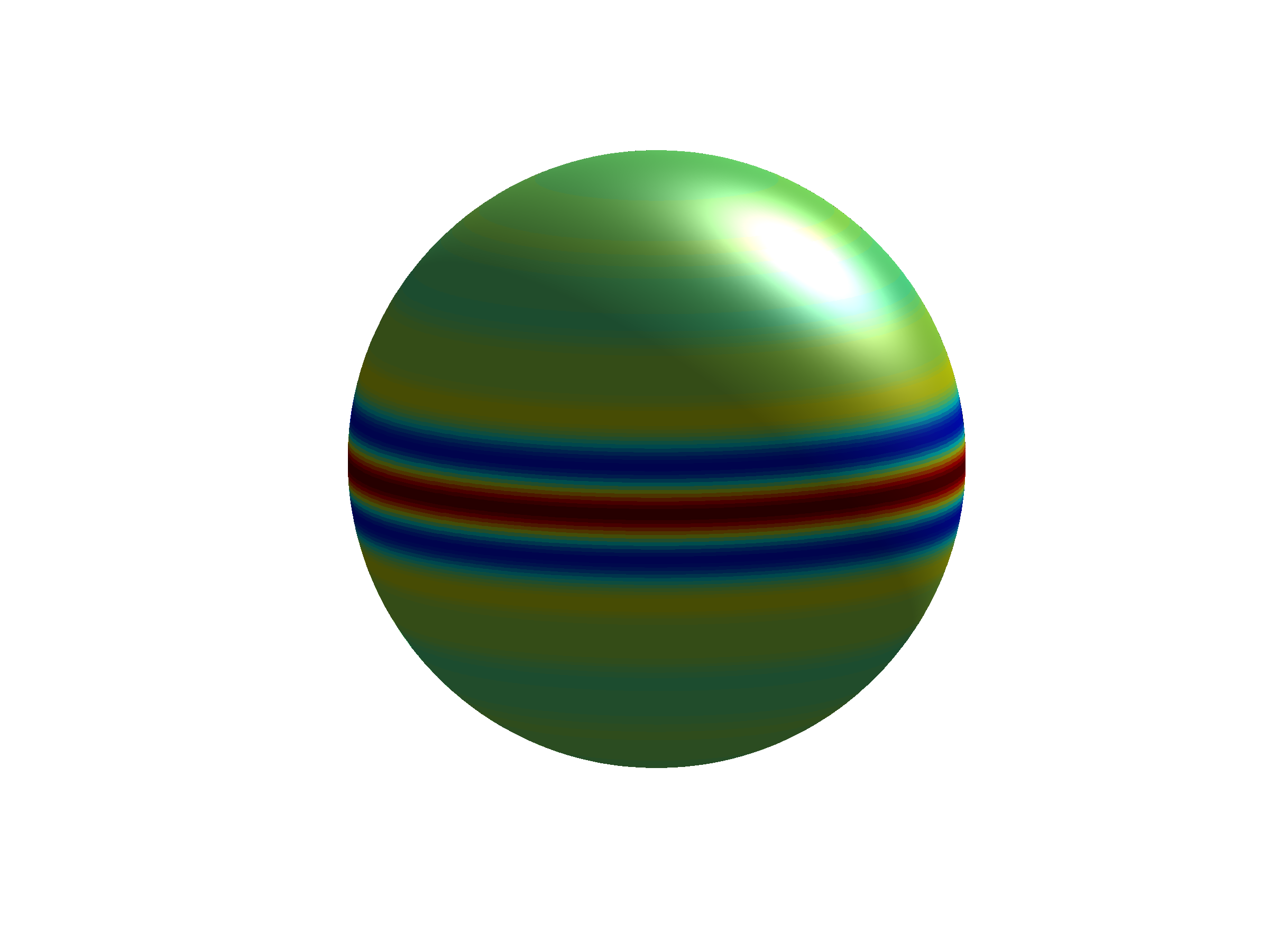}}\\
  \caption{Spherical ridgelets, with axis aligned with the North pole,
    for various wavelet scales. Notice that the constructed ridgelets are constant
    along ridges defined by great circles and wavelets transverse to
    ridges.}
  \label{fig:ridgelets}
\end{figure}

The rotated ridgelet
$\rotarg{(\sas)} \srwav^{(\wscale)}(\saa\p,\sab\p)$ should be constant
along the great circle defined by
$\vect{\hat{\sa}} \cdot \vect{\hat{\sa}\p} = 0$ and a wavelet
transverse to the ridge defined by the great circle.
Such a ridgelet on the sphere can be constructed from an axisymmetric
convolution of the Funk-Radon kernel $\sradondelta$ with the
axisymmetric wavelet ${}_0 \wav^{(\wscale)}$:
\begin{equation}
  \srwav^{(\wscale)}(\sas)
  \equiv
  (\sradondelta \convaxisym {}_0 \wav^{(\wscale)})(\sas)
  \spcend.
\end{equation}
In \fig{\ref{fig:ridgelets}} ridgelets are plotted for various scales
$\wscale$.  Notice that the ridgelets exhibit precisely the structure
desired -- probing signal content
along great circles (\cf\ global lines).

The ridgelet transform of \eqn{\ref{eqn:ridgelet_analysis}} can then
be viewed as the composition of a spherical Radon transform followed by a wavelet transform:
\begin{align}
  \rwcoeff & {}^{\srwav^{(\wscale)}} (\sas)
  \equiv
  (\rwavop^{\srwav^{(\wscale)}} \fs)(\sas)
  \equiv
  ( \fs \convaxisym \srwav^{(\wscale)}) (\sas) \nonumber \\
  &= ( \fs \convaxisym \sradondelta \convaxisym {}_0 \wav^{(\wscale)}) (\sas)
  \spcend .
\end{align}

A ridgelet scaling function $\srwavs^{(\wscale)} \in \ltwo(\sphere)$
must be defined to capture the low-frequency content of the
signal analysed:
\begin{equation}
  \srwavs^{(\wscale)}(\sas)
  \equiv
  (\sradondelta \convaxisym {}_0 \wavs^{(\wscale)})(\sas)
  \spcend.
\end{equation}
In terms of operators these relations can be written as,
\begin{equation}
  \rwavop^{\srwav^{(\wscale)}} = \wavop^{{}_0\wav^{(\wscale)}} \sradon
  \spcend \quad \text{and} \quad \rwavop^{\srwavs^{(\wscale)}} = \wavop^{{}_0\wavs^{(\wscale)}} \sradon
  \spcend.
\end{equation}
We write the ridgelet transform for all ridgelets and the ridgelet
scaling function by
\begin{equation}
  \label{eqn:ridgelet_analysis_operator}
  \vect{\rwcoeff} (\sas)
  \equiv
  (\vect{\rwavop} \: \fs)(\sas)
  =
  ({}_0 \vect{\wavop} \: \vect{\sradon} \: \fs)(\sas)
  \spcend ,
\end{equation}
where bold notation represents a collection of coefficients.
\par
For antipodal signals \fs\ can be synthesised exactly from its ridgelet
coefficients simply by:
\begin{equation}
  \label{eqn:ridgelet_synthesis_operator}
  \fs(\sas)
  =
  (\vect{\sradon}^{-1} {}_0 \vect{\wavop}^{-1} \vect{\rwcoeff}) (\sas)
  \spcend .
\end{equation}

\section{Evaluation}
\label{sec:evaluation}

Our spherical Radon and ridgelet transforms have been added to the
existing \stwoletcode\ \cite{leistedt:s2let_axisym, mcewen:s2let_spin}
code that supports the exact and efficient computation of
scale-discretised wavelet transforms on the sphere, which is
publicly available\footnote{\url{http://www.s2let.org}}, and relies on
the \sshtcode\footnote{\url{http://www.spinsht.org}} code
\cite{mcewen:fssht} to compute spherical harmonic transforms and the
\fftwcode\footnote{\url{http://www.fftw.org}} code to compute Fourier
transforms.  In this section we evaluate, on simulations of random
antipodal signals on the sphere, the numerical accuracy, computation
time and asymptotic scaling of the \stwoletcode\ implementation of the
ridgelet transform on the sphere.

\subsection{Simulations}

We simulate band-limited test signals on the sphere defined by
uniformly random spherical harmonic coefficients
$\sshc{\f}{\el}{\m}{\spin}$ with real and imaginary components in $[-1,1]$.  For $\el+\spin$ odd we set harmonic
coefficients to zero to satisfy the antipodal symmetry condition required for
inversion. We then compute
an inverse spherical harmonic transform to recover a band-limited
signal on the sphere.  A forward spherical ridgelet transform is then
performed, followed by an inverse transform to synthesise the original
signal from its ridgelet coefficients.  Ten simulated signals are
considered for range of band-limits $\elmax$ are considered (band-limits of
at least $\elmax=4096$ are feasible; \cf\ \cite{mcewen:fssht}).  All numerical
experiments are performed on a 2011 Macbook Air, with a 1.8$\:$GHz
Intel Core i7 processor and 4$\:$GB of RAM. Note that all numerical and
computational results are identical when considering spin signals.

\subsection{Numerical accuracy}

Numerical accuracy is quantified by the maximum absolute error between the spherical harmonic
coefficients of the original test signal $\fslm^{\rm o}$ and the
recomputed values $\fslm^{\rm r}$, \ie\
$
  \epsilon = \mathop{\rm max}_{\el,\m} \:
  \bigl | \fslm^{\rm r} - \fslm^{\rm o} \bigr |
$.
Results of the numerical accuracy tests, averaged over ten random test
signals, are plotted in \fig{\ref{fig:accuracy}}. The numerical accuracy of the round-trip
transform is close to machine precision and found empirically to scale
as $\order(\elmax^2)$, with a factor of $\order(\elmax)$ coming from both the inverse
Radon and wavelet transforms.

\subsection{Computation time}

\begin{figure}[t]
  \centering
  \subfigure[Maximum error \label{fig:accuracy}]{\includegraphics[width=.47\linewidth]{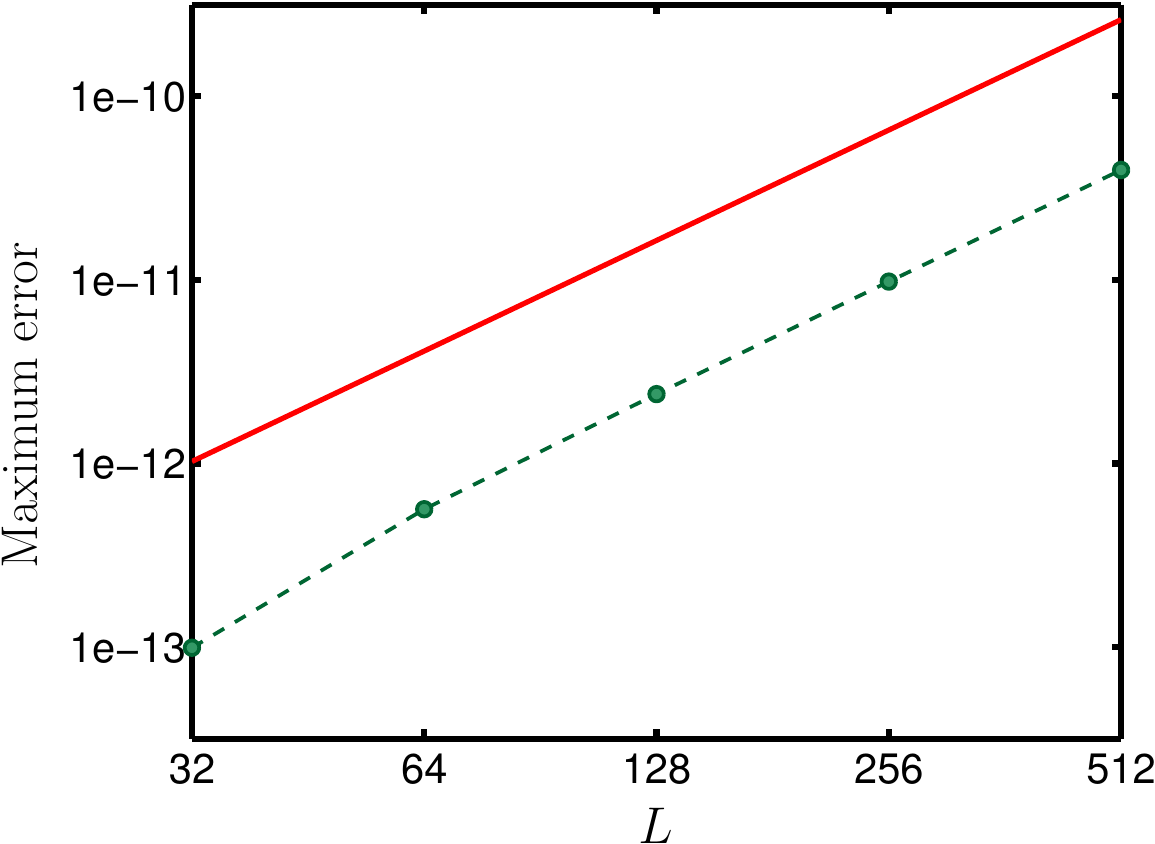}}
  \subfigure[Computation time \label{fig:timing}]{\includegraphics[width=.47\linewidth]{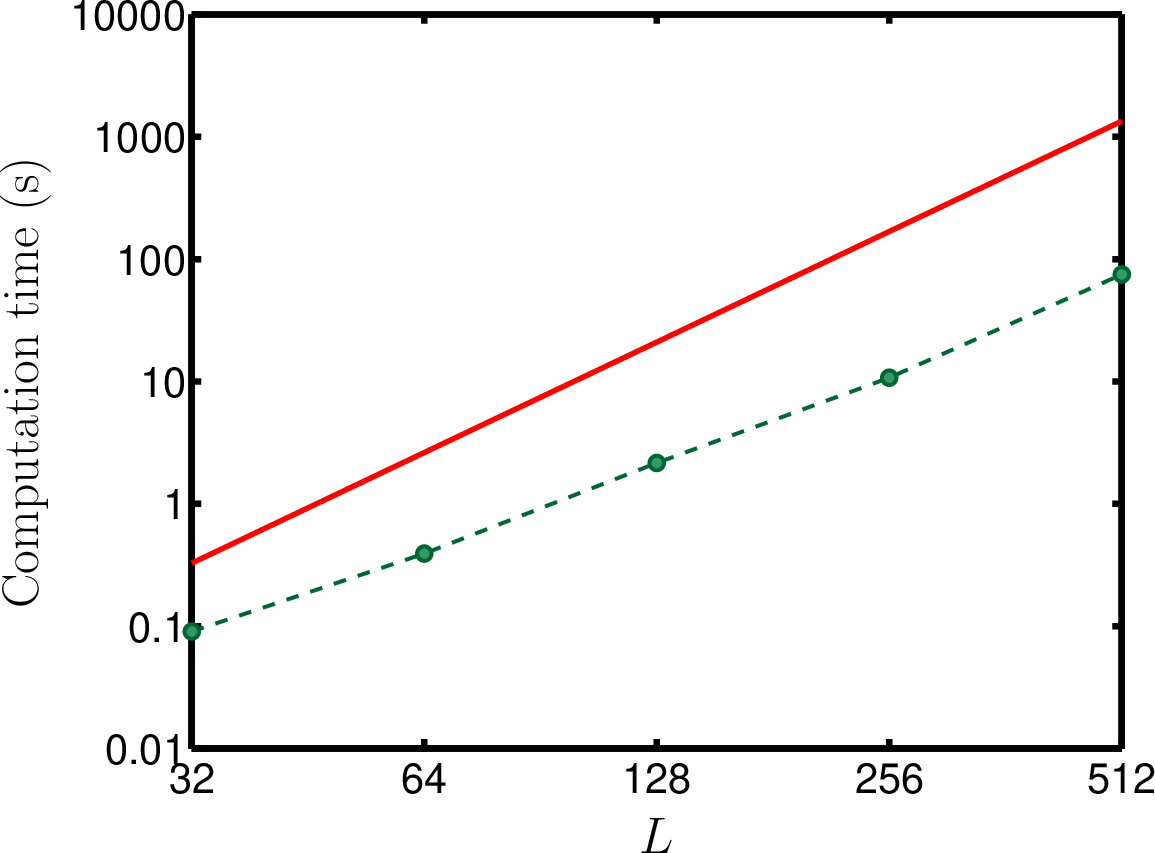}}
  \caption{Numerical accuracy and computation time of the spherical
    ridgelet transform, averaged over ten round-trip transforms of
    random test signals.  Numerical accuracy close to machine
    precision is achieved and found empirically to scale as
    $\order(\elmax^2)$, with a factor of $\order(\elmax)$ coming from
    the inversion of each of the spherical Radon and spherical wavelet
    transforms.  Computation time is found empirically to scale as
    $\order(\elmax^3)$, as expected theoretically.  $\order(\elmax^2)$
    and $\order(\elmax^3)$ scaling is shown by the solid red lines in
    panels (a) and (b) respectively.}
\end{figure}

Computation time is quantified by the time taken
to perform a forward and inverse spherical ridgelet transform.
Results of the computation time tests, averaged over ten random test
signals, are plotted in \fig{\ref{fig:timing}}. The computational complexity of
the ridgelet transform is dominated by the spherical harmonic
transform, which scales as $\order(\elmax^3)$, as seen in \fig{\ref{fig:timing}}.

\section{Illustration}
\label{sec:illustration}

In this section we illustrate the application of the spherical
ridgelet transform to the analysis of diffusion MRI signals acquired
on the sphere.

\subsection{Diffusion MRI  signals on the sphere}
\label{sec:illustration:hardi}

Diffusion MRI can be used to study neuronal connections
by measuring the diffusion of water molecules along white matter
fibers. In so-called high angular resolution diffusion imaging
(HARDI), diffusion MRI signals are sampled on spherical shells in
each voxel of the brain.  The orientation distribution function (ODF) is
approximately given by the
spherical Radon transform of the HARDI signal acquired over a single spherical shell
\cite{tuch:2004}.  Often acquired data is noisy and
incomplete, motivating the development of reqularized ODF recovery
techniques (for a review see \cite{alexander:2005}).

The HARDI signal is modelled by a sum of weighted Gaussians, where
each Gaussian corresponds to a different fiber passing through the voxel,
and is given by (\eg\ \cite{michailovich:2010a})
\begin{equation}
  S(\vect{\hat{\sa}}) = \sum_i p_i
  \exp{- b \vect{\hat{\sa}}^{\rm T} \mathsf{D}_i \vect{\hat{\sa}}}
  \spcend ,
\end{equation}
where $\mathsf{D}_i$ is the $3\times3$ diffusion tensor corresponding
to fiber $i$, $b$ is an acquisition configuration constant,
and $p_i$ are fiber weights.  We adopt the same parameters as the in silico experiments of
\cite{michailovich:2010a}. Three fibers are considered, with
$\mathsf{D}_i$ computed from $\mathsf{D}$ by random rotations.
The simulated HARDI signal and the corresponding ODF are plotted in \fig{\ref{fig:hardi_coefficients}}.

\subsection{Diffusion MRI spherical ridgelet decomposition}
\label{sec:illustration:ridgelet}

Since the diffusion MRI HARDI signal is composed of a sum of
contributions for each fibre that have their energy concentrated along
great circles, it is suggested in
\cite{michailovich:2010a,michailovich:2010b} that spherical ridgelets,
which have their energy similarly distributed, are effective for
representing HARDI signals and, in particular, more suitable than spherical
wavelets. We demonstrate and validate these predictions by examining
a HARDI signal in both spherical wavelet and ridgelet representations.

\begin{figure}
  \centering
  \subfigure[HARDI signal \label{fig:hardi:hardi}]{\includegraphics[trim=50mm 20mm 45mm 25mm,clip=true,width=.35\linewidth]{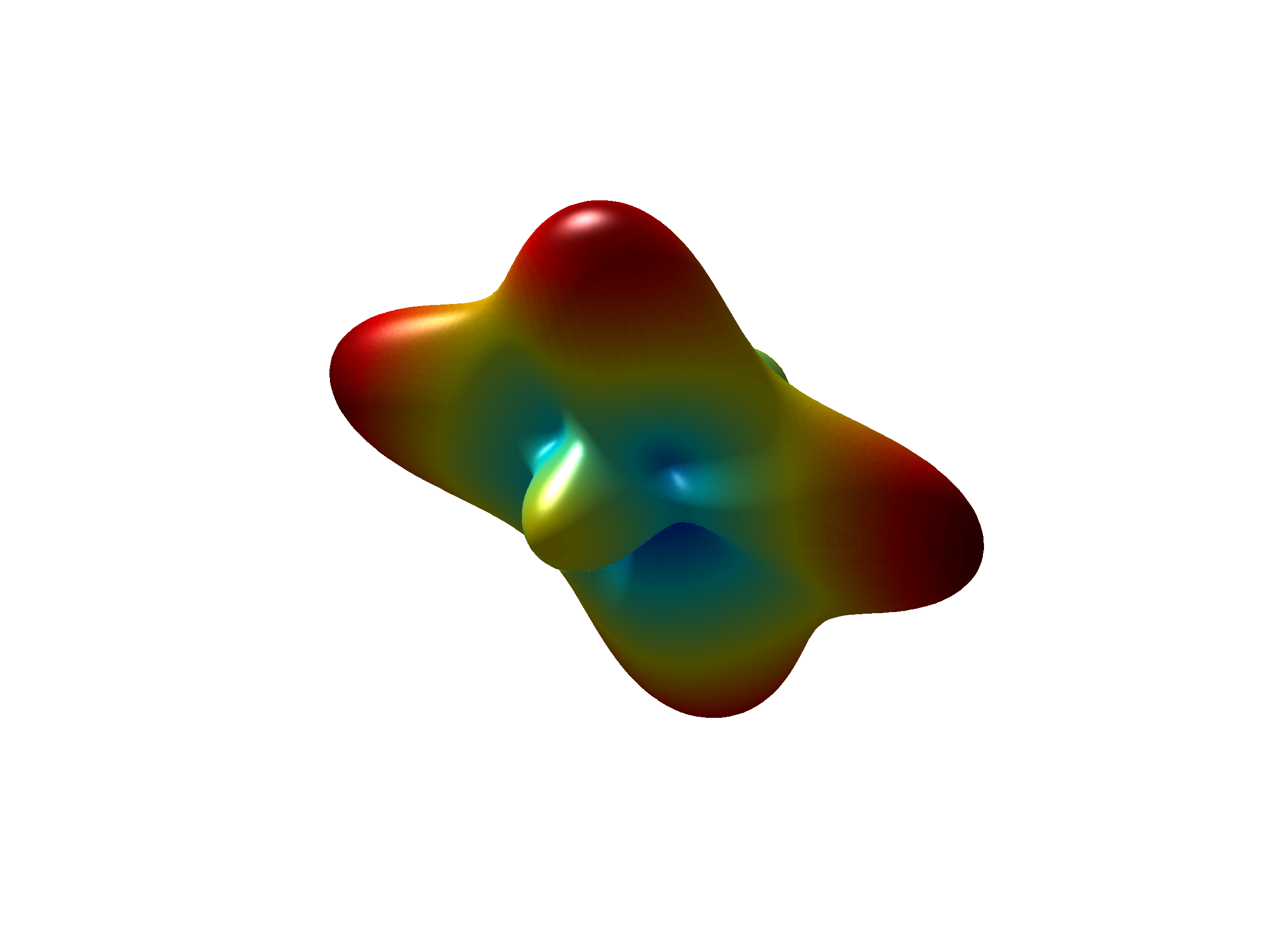}}\quad
  \subfigure[ODF signal \label{fig:hardi:odf}]{\includegraphics[trim=50mm 18mm 50mm 13mm,clip=true,width=.3\linewidth]{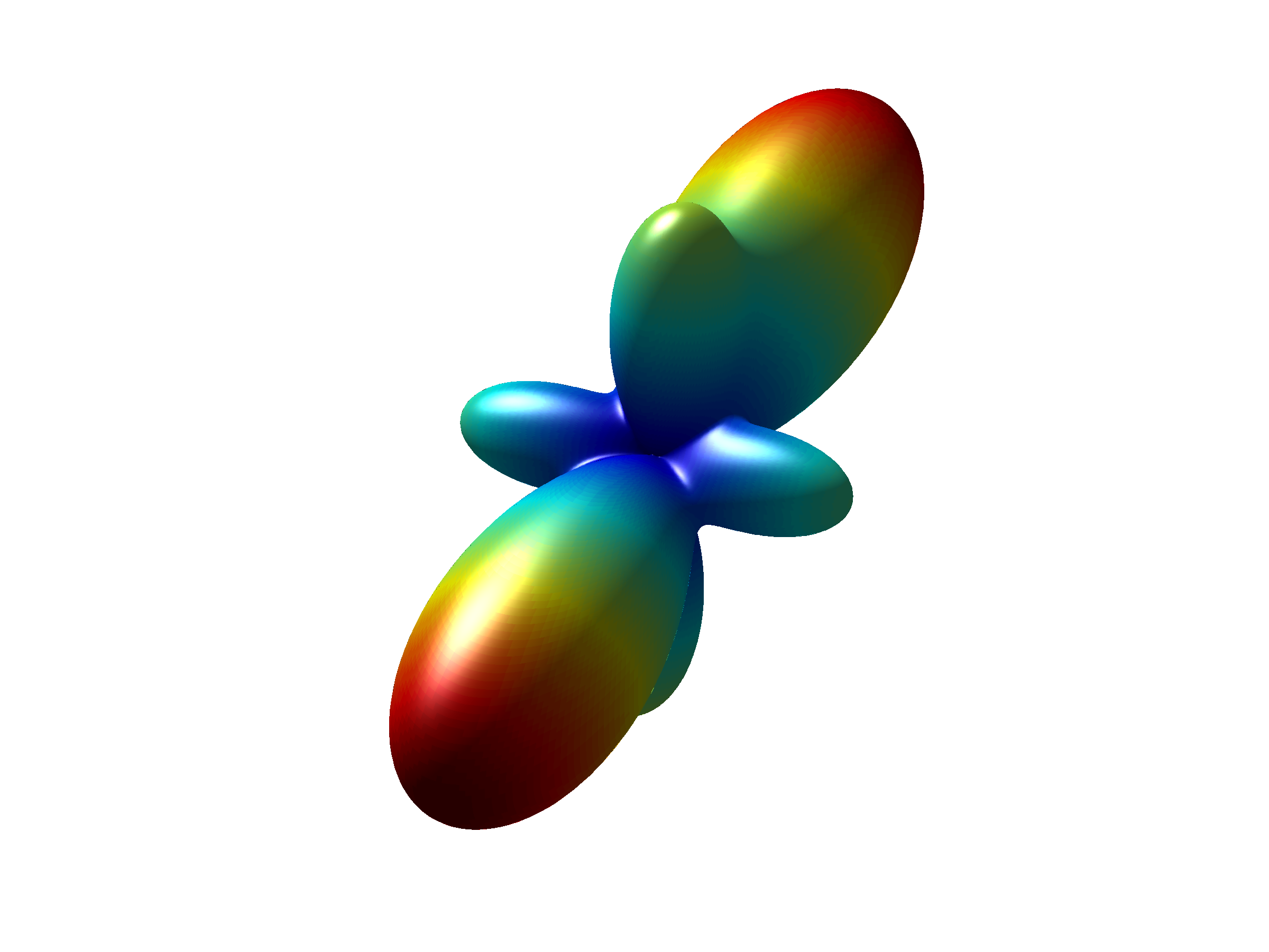}}\\
  \subfigure[Wavelet coefficients for $\wscale=5$]{\includegraphics[trim=50mm 30mm 50mm 26mm,clip=true,width=.45\linewidth]{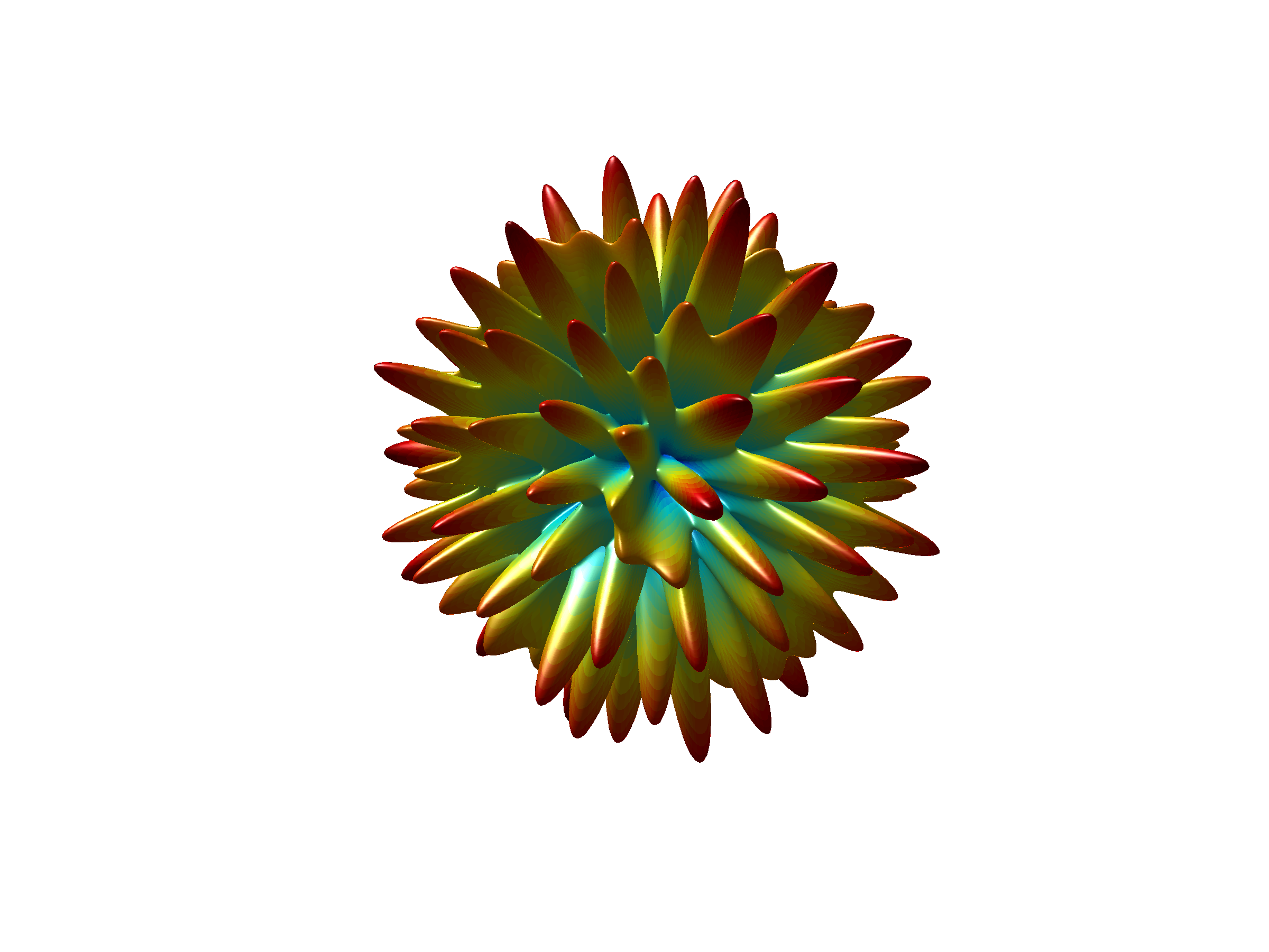}}\quad
  \subfigure[Ridgelet coefficients for $\wscale=5$]{\includegraphics[trim=50mm 30mm 50mm 26mm,clip=true,width=.45\linewidth]{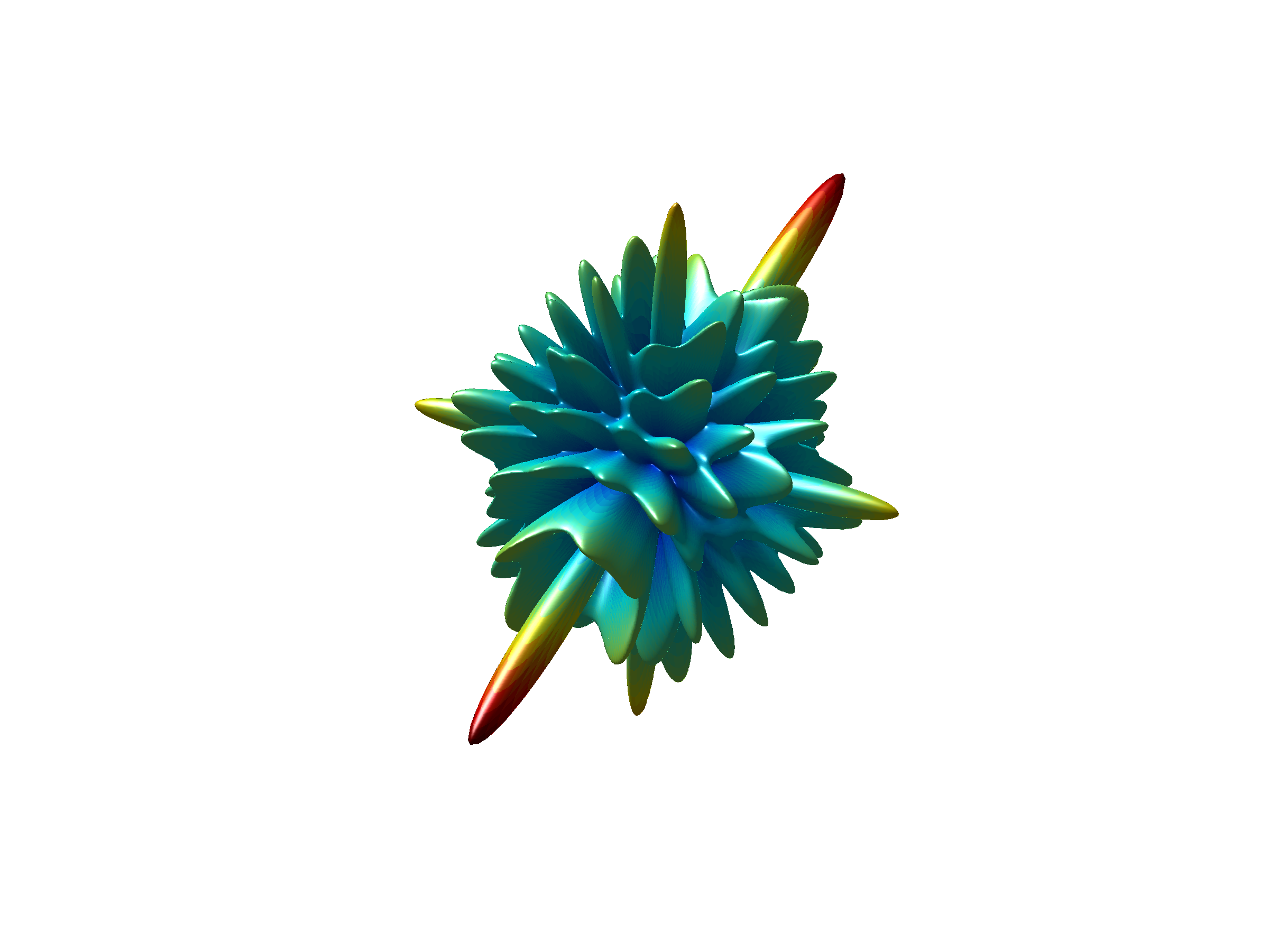}}\\
  \caption{Parametric plots of spherical wavelet (left column, bottom) and
    ridgelet (right column, bottom) coefficients of the HARDI signal plotted
    in the top row.  Notice that ridgelet coefficients
    are more sparse (\ie\ fewer large coefficients) than the wavelet
    coefficients.}
  \label{fig:hardi_coefficients}
\end{figure}

\begin{figure}
  \centering
  \includegraphics[width=.8\linewidth]{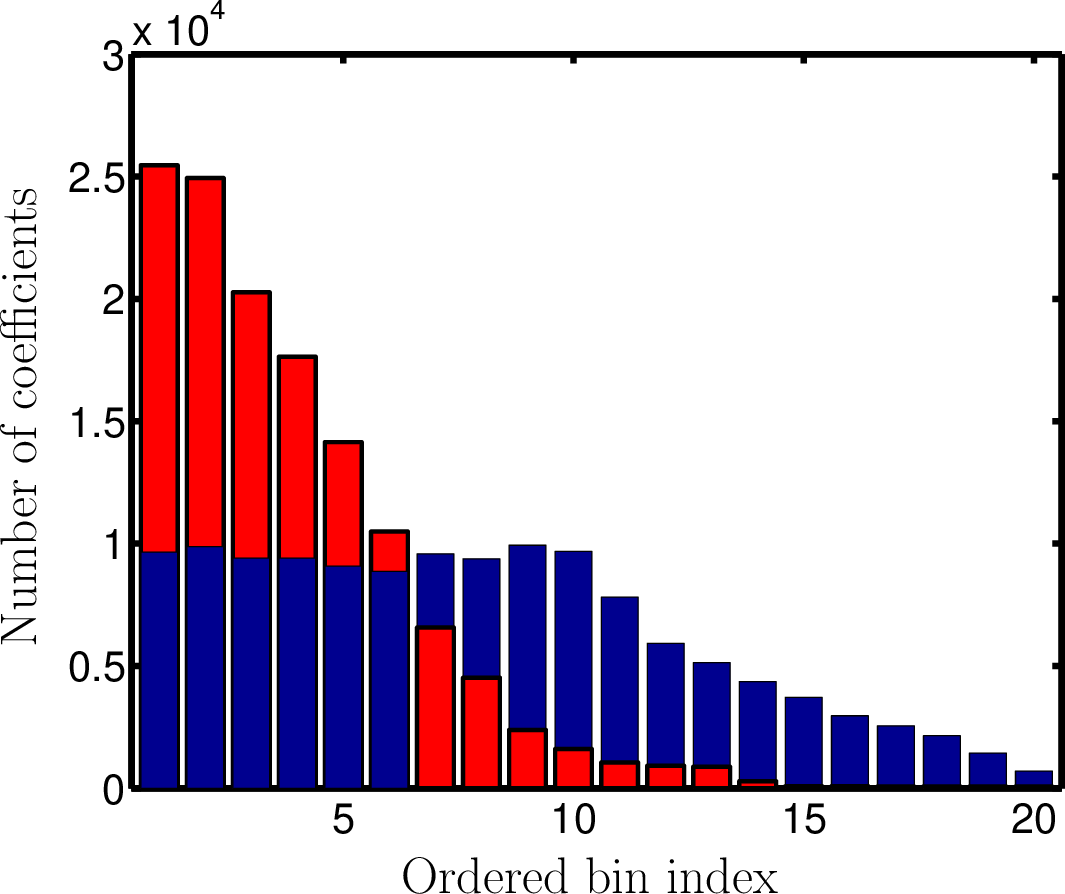}
  \caption{Histogram of (the absolute value of) wavelet (blue) and ridgelet
    (red) coefficients for scale $\wscale=4$ of the HARDI signal
    plotted in \fig{\ref{fig:hardi:hardi}}.  Notice that ridgelet
    coefficients are sparser than wavelet coefficients, with the
    ridgelet coefficients containing many coefficients close to zero
    and fewer large coefficients.  The sparseness of the ridgelet
    coefficients of the HARDI signal demonstrates the suitability of
    spherical ridgelets for diffusion MRI.}
  \label{fig:dmri_histogram}
\end{figure}

In \fig{\ref{fig:hardi_coefficients}} we plot wavelet and ridgelet
coefficients of the HARDI signal simulated in
\sectn{\ref{sec:illustration:hardi}} for a range of scales $\wscale$.
It is clear that ridgelet coefficients of the HARDI signal are sparser
than wavelet coefficients, which exhibit many large peaks.  For the
ridgelet decompositions (\fig{\ref{fig:hardi_coefficients}}, right
column), the dominant directions of the ODF signal
(\fig{\ref{fig:hardi:odf}}) are visible by eye, which is not the case
for the wavelet decompositions (\fig{\ref{fig:hardi_coefficients}}, left
column).
In \fig{\ref{fig:dmri_histogram}} we plot histograms of wavelet and
ridgelet coefficients for scale $\wscale=4$.
The sparseness of HARDI signals in the spherical ridgelet
decomposition, as demonstrated in this simple illustration, can be
exploited in practical applications to handle noisy and incomplete
data.

\section{Conclusions}
\label{sec:conclusions}

The publicly available ridgelet transform presented in this letter
is defined natively on the sphere, probes signal content globally
along great circles, does not exhibit blocking artefacts, supports spin signals, and exhibits an explicit inverse
transform.

We present a novel take on the spherical Radon transform, viewing it as
a convolution with an axisymmetric kernel. Such a representation leads to
a straightforward derivation of the harmonic action of the spherical Radon transform,
which motives an exact inversion technique for signals that exhibit
antipodal symmetry.  Consequently, our spherical ridgelet transform
also permits the exact inversion for antipodal signals.

We demonstrate that the numerical accuracy of our transforms is close to
machine precision and can be applied to large data-sets supporting high band-limits
$\elmax$, with computational complexity scaling as $\order(\elmax^3)$.
Finally, we illustrate the effectiveness of spherical ridgelets for
imaging white matter fibers in the brain by diffusion MRI.

\bibliographystyle{IEEEtran}
\bibliography{bib_myname,bib_journal_names_short,bib}




%








\end{document}

%% file: macros.tex
\newcommand{\eqn}[1]{(#1)}

\newcommand{\fig}[1]{Fig.~#1}

\newcommand{\sectn}[1]{Sec.~#1}

\newcommand{\eg}{\mbox{\it e.g.}}
\newcommand{\ie}{\mbox{\it i.e.}}

\newcommand{\cf}{\mbox{\it cf.}}

\newcommand{\healpix}{{\tt HEALPix}}

\newcommand{\sshtcode}{{\tt SSHT}}

\newcommand{\stwoletcode}{{\tt S2LET}}

\newcommand{\spcend}{\ensuremath{\:}}

\newcommand{\cconj}{\ensuremath{\ast}}

\newcommand{\ltwo}{\ensuremath{\mathrm{L}^2}}
\newcommand{\sphere}{\ensuremath{{\mathbb{S}^2}}}
\newcommand{\sothree}{\ensuremath{{\mathrm{SO}(3)}}}

\newcommand{\vect}[1]{\ensuremath{\mbox{\boldmath ${#1}$}}}

\newcommand{\dx}{\ensuremath{\mathrm{\,d}}}
\newcommand{\dmu}[1]{\ensuremath{\dx \Omega(#1)}}

\newcommand{\innerp}[2]{\ensuremath{\langle {#1},\: {#2} \rangle}}

\newcommand{\sa}{\ensuremath{\omega}}
\newcommand{\saa}{\ensuremath{\theta}}
\newcommand{\sab}{\ensuremath{\varphi}}
\newcommand{\sas}{\ensuremath{\saa, \sab}}

\newcommand{\euls}{\ensuremath{\eula, \eulb, \eulc}}
\newcommand{\eula}{\ensuremath{\alpha}}
\newcommand{\eulb}{\ensuremath{\beta}}
\newcommand{\eulc}{\ensuremath{\gamma}}

\newcommand{\el}{\ensuremath{\ell}}
\newcommand{\m}{\ensuremath{m}}

\newcommand{\spin}{\ensuremath{s}}

\newcommand{\elmax}{\ensuremath{{L}}}

\newcommand{\p}{\ensuremath{^\prime}}

\newcommand{\kron}[2]{\ensuremath{\delta_{{#1}{#2}}}}

\renewcommand{\exp}[1]{\ensuremath{{\rm e}^{#1}}}
\newcommand{\shfarg}[3]{\ensuremath{Y_{#1#2}({#3})}}

\newcommand{\shc}[3]{\ensuremath{{#1}_{{#2}{#3}}}}

\newcommand{\sshf}[3]{\ensuremath{{{}_{#3} Y_{#1#2}}}}

\newcommand{\sshc}[4]{\ensuremath{{}_{#4} {#1}_{{#2}{#3}}}}
\newcommand{\sshcc}[4]{\ensuremath{{}_{#4} {#1}_{{#2}{#3}}^\cconj}}

\newcommand{\aleg}[3]{\ensuremath{P_{#1}^{#2}({#3})}}

\newcommand{\rotarg}[1]{\ensuremath{\mathcal{R}_{#1}}}

\newcommand{\f}{\ensuremath{f}}
\newcommand{\fs}{\ensuremath{{}_\spin f}}

\newcommand{\fslm}{\ensuremath{\shc{\fs}{\el}{\m}}}

\newcommand{\wavop}{\ensuremath{\mathcal{W}}}
\newcommand{\wav}{\ensuremath{\psi}}

\newcommand{\wavs}{\ensuremath{\Phi}}

\newcommand{\wscale}{\ensuremath{j}}

\newcommand{\sradon}{\ensuremath{\mathcal{S}}}
\newcommand{\sradondelta}{\ensuremath{\xi}}

\newcommand{\srwav}{\ensuremath{{}_\spin\psi}}

\newcommand{\srwavs}{\ensuremath{{}_\spin\phi}}

\newcommand{\rwcoeff}{\ensuremath{G}}

\newcommand{\rwavop}{\ensuremath{\mathcal{G}}}

\newcommand{\sumlm}{\ensuremath{\sum_{\el=0}^{\infty} \sum_{\m=-\el}^\el}}

\newcommand{\order}{\ensuremath{\mathcal{O}}}

\newcommand{\convaxisym}{\ensuremath{\odot}}

